\documentclass[aps,twocolumn,pra,showpacs,10pt]{revtex4-1}

\usepackage{epsfig,amssymb,amsmath,graphicx}
\usepackage{hyperref}
\newcommand{\sinc}{{\rm sinc}}

\begin{document}

\title{Quantum imaging of spin states in optical lattices}
\author{James Douglas$^1$ and Keith Burnett$^2$}
\affiliation{$^1$Clarendon Laboratory, University of Oxford, Oxford OX1 3PU, United Kingdom
\\$^2$University of Sheffield, Western Bank, Sheffield S10 2TN, United Kingdom}
\begin{abstract}
We investigate imaging of the spatial spin distribution of atoms in optical lattices using non-resonant light scattering. We demonstrate how scattering spatially correlated light from the atoms can result in spin state images with enhanced spatial resolution. Furthermore, we show how using spatially correlated light can lead to direct measurement of the spatial correlations of the atomic spin distribution.
\end{abstract}
\date{12 July 2010}
\pacs{37.10.Jk,42.50.Ct,42.50.Hz,42.30.Va}
\maketitle

%%%%%%%%%%%%%%%%%%%%%%%%%%%%%%%%%%%%%%%%%%%%%%%%%%%%%%%%%%%%%%%%%%%%%%%%%%%%%%%%%%%%%%%%%%%%%%%%%%%%%%%%%%%%%%%%%%%%%%%%%%%%%%%%%%%%%%%%%%%%%%%%%%%%%%%%%%%%%%%%%%%%%%%%%%%%%%%%%%%%%%%%%%%%%%%%%%%%%%%%%%%%%%%%%%%%%%%%%%%%%%%%%%%%%%%%%%%%%%%%%%%%%%%%%%%%%%%%

Experiments with ultracold atoms in optical lattices have become a integral part of investigations into the fundamentals of quantum mechanics,  precision measurement and quantum computing, and have become a valuable tool for the simulation of condensed matter problems \cite{Bloch2008a}. These experiments typically use time of flight absorption imaging \cite{Greiner2002a}, or more recently noise correlation imaging \cite{Altman2004a} to examine the atomic state. These methods destructively probe the momentum state of the atomic cloud with resonant light after releasing the atoms from the lattice and allowing them to expand.

As experiments with optical lattices become more advanced, atomic spin has become an important variable \cite{Lewenstein2007a}, and new ways to probe it must be investigated \cite{Cherng2007a,Vega2008a,Eckert2008a,Roscilde2009a}.
Here we consider imaging the spatial atomic spin distribution non-destructively while the atoms remain \textit{in situ}.  In particular we look at off-resonant light scattered from the atoms, which is then collected by a microscope to form an image on a detector. We consider incoming light from coherent beams and also from spatially correlated beams. Where in the later case we build on the ideas developed in the field of quantum imaging \cite{Giovannetti2009a,Lugiato2002a} to show how the resolution of images can be enhanced and how spatial correlation functions of the atomic spin distribution can be measured. These concepts become increasing useful for optical lattices where the lattice spacing is often similar to the wavelength of the probe light and coherent imaging becomes less useful due to the diffraction limit.

Our approach is organized as follows. We begin by describing the interaction between atomic spins and off-resonant light that our imaging proposal is based on. We then develop a description of how multi-photon light states are imaged following their interaction with the atoms. We then apply our theory to imaging of a number of possible spin states. Finally we show how using spatially correlated imaging light can also lead to direct measurement of spatial correlation functions.

%%%%%%%%%%%%%%%%%%%%%%%%%%%%%%%%%%%%%%%%%%%%%%%%%%%%%%%%%%%%%%%%%%%%%%%%%%%%%%%%%%%%%%%%%%%%%%%%%%%%%%%%%%%%%%%%%%%%%%%%%%%%%%%%%%%%%%%%%%%%%%%%%%%%%%%%%%%%%%%%%%%%%%%%%%%%%%%%%%%%%%%%%%%%%%%%%%%%%%%%%%%%%%%%%%%%%%%%%%%%%%%%%%%%%%%%%%%%%%%%%%%%%%%%%%%%%%%%

\section{Interaction between light and atomic spins}

Many ultracold atom experiments are done with alkali atoms and we thus restrict our attention to the interaction between alkali atoms in their ground electronic state and detuned light. In the off-resonant limit, i.e., where the detuning from the $S_{1/2} \rightarrow P_{1/2},P_{3/2}$ transitions is larger than the natural line widths and also larger than
the Rabi frequency of any particular radiation mode, the excited states can be adiabatically eliminated from the interaction Hamiltonian. In the dipole and rotating wave approximations the interaction between the radiation and atomic fields is then given by
the effective Hamiltonian
\begin{equation}
\hat{H}
=\sum_{m_1,m_2}\int\! d\mathbf{r}\hat{\Psi}^\dagger_{m_2}(\mathbf{r})
(\mathbf{\tilde{E}}^-(\mathbf{r})\tensor{\alpha}_{m_1,m_2}\mathbf{\tilde{E}}^+(\mathbf{r}))\hat{\Psi}_{m_1}(\mathbf{r}),
\end{equation}
where $\tensor{\alpha}_{m_1,m_2}$ is the polarizability tensor  \cite{Hammerer2010a,Vega2008a,Deutsch1998a}. Here the atomic field operators $\hat{\Psi}^\dagger_{m}(\mathbf{r})$ and $\hat{\Psi}_{m}(\mathbf{r})$ create and destroy atoms in the ground hyperfine state $|F m\rangle$, and $\mathbf{\tilde{E}}^+(\mathbf{r})$ and $\mathbf{\tilde{E}}^-(\mathbf{r})$ are the slowly varying positive and negative frequency components of the electric field.

As discussed in Hammerer \textit{et al.} \cite{Hammerer2010a},\ the polarizability tensor can be decomposed into three irreducible components of rank zero, one and two. The rank zero term leads to a spin-independent interaction, which does not play a part in the imaging interactions we consider in this work. The rank one term is responsible for the Faraday effect and is the interaction we use for imaging. The rank two term also leads to a spin dependent interaction, but is typically at least an order of magnitude smaller than the second term and asymptotically goes to zero as the detuning increases, and we neglect it here.

Keeping only the rank one component the interaction becomes
\begin{equation}
\hat{H}
=a_1(\Delta)\sum_{m_1,m_2}\int\! d\mathbf{r}\boldsymbol{\hat{\rho}}(\mathbf{r})\cdot(\mathbf{\tilde{E}}^-(\mathbf{r})\times\mathbf{\tilde{E}}^+(\mathbf{r})).
\end{equation}
Here $\boldsymbol{\hat{\rho}}(\mathbf{r})=\sum_{m_1,m_2} \hat{\Psi}^\dagger_{m_2}(\mathbf{r})\langle F m_2|\hat{\mathbf{F}}/\hbar|F m_1\rangle\hat{\Psi}_{m_1}(\mathbf{r})$ is a vector operating on the atomic spin and $a_1(\Delta)$ is the detuning dependent coupling constant (see \cite{Hammerer2010a}).

%%%%%%%%%%%%% FIGURE 1 %%%%%%%%%%%%%%%%%%%%%%%%%%%%%%%

\begin{figure}
\begin{center}
\includegraphics{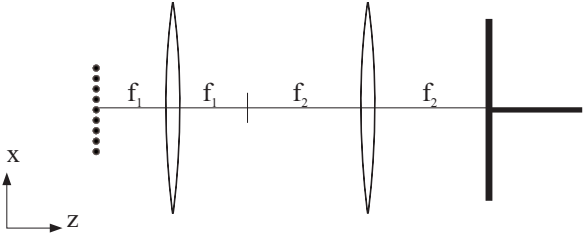}
\end{center}
\caption{Imaging scheme. Light propagating in the z direction interacts with atoms in an optical lattice, then passes through a diffraction limited two lens system and is detected at the focal plane of the second lens.}
\label{imaging_setup}
\end{figure}

%%%%%%%%%%%%% FIGURE 1 %%%%%%%%%%%%%%%%%%%%%%%%%%%%%%%

This effective interaction Hamiltonian describes light scattering from the incoming beam, where photons scatter from the atoms in the lattice and take away information about the spin distribution. These photons can then be imaged onto a detector as shown in Figure \ref{imaging_setup}. To make calculations for this system practical however, the full three dimensional, multimode interaction Hamiltonian must first be simplified. In considering this type of imaging our primary concern is the spatial description of the light as it interacts with the atoms in the optical lattice and propagates through the system. With this in mind, we assume the photons in the light field are approximately monochromatic with frequency $\omega$. We also consider one-dimensional optical lattices in the $x$-direction, with the result that the useful variation in an image is also in the $x$-direction. The system can then be approximately described by a light field that only varies in two dimensions, $x$ and $z$. 
Numerical calculations of images for three dimensional systems show that this simplified theory captures the nature of the imaging process well, and provides a good model for images integrated over the $y$-direction.

Tsang \cite{Tsang2007a} describes the quantization of a two dimensional electromagnetic field in the monochromatic approximation. We generalize Tsang's formalism to include both possible polarizations, in which case the electric field operator becomes
\begin{equation}
\tilde{\mathbf{E}}^+(x,z) = i \sum_\sigma\sqrt{\frac{\eta}{2 \pi}}\int_{-k}^{k}\!\!\!\!d \kappa \gamma(\kappa) \hat{a}_\sigma(\kappa)\boldsymbol\epsilon_\sigma(\kappa)e^{i \kappa x +  i\sqrt{k^2-\kappa^2} z},
\end{equation}
where  $\eta = \hbar k/(2 \epsilon_0  L_y T)$, $\gamma(\kappa) = (1-\kappa^2/k^2)^{-1/4}$, $k= \omega/c = 2\pi/\lambda$, and $T$ and $L_y$ are the normalization time scale and y-dimension length scale respectively (see \cite{Tsang2007a} for further details). 

The operators $\hat{a}_1(\kappa)$ and $\hat{a}_2(\kappa)$ destroy photons with wavevector $\kappa \mathbf{\hat{x}} + k \mathbf{\hat{z}}/\gamma(\kappa)^2$ and polarization vectors $\boldsymbol\epsilon_1(\kappa)=\hat{\mathbf{x}}/\gamma(\kappa)^2-\kappa\hat{\mathbf{z}}/k$ and $\boldsymbol\epsilon_2(\kappa)=\hat{\mathbf{y}}$ respectively. They obey the commutator relation $[ \hat{a}_\sigma(\kappa), \hat{a}_{\sigma'}(\kappa')] = \delta(\kappa-\kappa')\delta_{\sigma,\sigma'}$. 
Using these operators we can define the $N$-photon momentum eigenstate
\begin{multline}
|\kappa_1,\ldots\kappa_n\rangle_1 \otimes|\kappa_{n+1},\ldots,\kappa_N\rangle_2 = \frac{1}{\sqrt{n!(N-n)!}}\\
\times\hat{a}_1^\dagger(\kappa_1)\ldots\hat{a}_1^\dagger(\kappa_n)\hat{a}_2^\dagger(\kappa_{n+1})\ldots
\hat{a}_2^\dagger(\kappa_N)|0\rangle,
\end{multline}
which has $n$ photons with $\boldsymbol{\epsilon}_1$-polarization and $N-n$ photons with $\boldsymbol{\epsilon}_2$-polarization.
These states can then be used as a basis to express a general state $|n\rangle_1|N-n\rangle_2$, which again has which has $n$ photons with $\boldsymbol{\epsilon}_1$-polarization and $N-n$ photons with $\boldsymbol{\epsilon}_2$-polarization,
\begin{multline}
|n\rangle_1|N-n\rangle_2 = \int^{k}_{-k}d\kappa_1\ldots d\kappa_N
\phi(\kappa_1,\ldots,\kappa_N)\\
\times|\kappa_1,\ldots\kappa_n\rangle_1 \otimes|\kappa_{n+1},\ldots,\kappa_N\rangle_2,
\end{multline}
where we refer to $\phi(\kappa_1,\ldots,\kappa_N)$ as the momentum distribution of the $N$-photon state.

We suppose the atoms are in an optical lattice in the $x$-direction with uniform site separation $a$ and expand the atomic field operators in terms of lowest band Wannier functions $\hat{\Psi}_{m}(\mathbf{r}) = \sum_j\hat{b}_{j,m} w(\mathbf{r}_j)$, which we assume are not dependent on the spin state of the atom.  The operator $\hat{b}_{j,m}$ destroys an atom with spin $m$ at lattice site $j$, centered at $\mathbf{r}_j = (x_j,0,0)$. We assume the lattice is strong enough so that the Wannier functions have negligible overlap between neighboring sites, in which case the spin operator becomes
\begin{equation}
\boldsymbol{\hat{\rho}}(\mathbf{r})=\sum_{j,m_1,m_2}\!\! |w(\mathbf{r}_j)|^2\hat{b}_{j,m_2}^\dagger\hat{b}_{j,m_1}\langle F m_2|\mathbf{\hat{F}}/\hbar|F m_1\rangle.
\end{equation}
Integrating over $y$ and $z$ then gives a vector operator dependent only on $x$
\begin{equation}
\boldsymbol{\hat{\rho}}(x) = \int dy dz\boldsymbol{\hat{\rho}}(\mathbf{r}).
\end{equation}

In terms of this spin operator and the two-dimensional electric field the effective interaction Hamiltonian becomes
\begin{widetext}
\begin{multline}
\hat{H}
= \frac{a_1(\Delta)\eta}{2\pi}\int dx\int_{-k}^{k}d \kappa_1 d\kappa_2 \gamma(\kappa_1)\gamma(\kappa_2)
e^{i (\kappa_2-\kappa_1)x}%\\\times
\boldsymbol{\hat{\rho}}(x)\cdot\left[
\left(\frac{\kappa_1}{k}\hat{a}^\dagger_1(\kappa_1)\hat{a}_2(\kappa_2)-\frac{\kappa_2}{k}\hat{a}^\dagger_2(\kappa_1)\hat{a}_1(\kappa_2)\right)\mathbf{\hat{x}}\right.\\\left.+
\left(\frac{\kappa_2}{k\gamma(\kappa_1)^2}-\frac{\kappa_1}{k\gamma(\kappa_2)^2}\right)
\hat{a}^\dagger_1(\kappa_1)\hat{a}_1(\kappa_2)\mathbf{\hat{y}}%\right.\\\left.
+\left(\frac{1}{\gamma(\kappa_1)^2}\hat{a}^\dagger_1(\kappa_1)\hat{a}_2(\kappa_2)-
\frac{1}{\gamma(\kappa_2)^2}\hat{a}^\dagger_2(\kappa_1)\hat{a}_1(\kappa_2)\right)\mathbf{\hat{z}}\right]
\label{monochrome_2d_h}
\end{multline}
\end{widetext}
In deriving Equation (\ref{monochrome_2d_h}) we used the approximation
\begin{equation}
\int\!dy dz |w(\mathbf{r}_j)|^2 e^{i(\sqrt{k^2-\kappa_2^2}-\sqrt{k^2-\kappa_1^2}) z} \approx \int dy dz|w(\mathbf{r}_j)|^2,
\end{equation}
which is very good for the imaging scenarios we consider, where the Wannier functions are confined in the $z$-direction tightly compared to the variation of the electric field in the $z$-direction.

Furthermore, for the imaging parameters we consider, the term proportional to 
$\boldsymbol{\hat{\rho}}(x)\cdot\mathbf{\hat{z}} = \hat{\rho}_z(x)$ in the effective Hamiltonian is much larger than the others and determines the dominant features of the images. In the following we neglect the other terms in our theoretical description. These terms lead to small changes that can be easily calculated and are included numerically in the calculations used to produce our figures.

%%%%%%%%%%%%%%%%%%%%%%%%%%%%%%%%%%%%%%%%%%%%%%%%%%%%%%%%%%%%%%%%%%%%%%%%%%%%%%%%%%%%%%%%%%%%%%%%%%%%%%%%%%%%%%%%%%%%%%%%%%%%%%%%%%%%%%%%%%%%%%%%%%%%%%%%%%%%%%%%%%%%%%%%%%%%%%%%%%%%%%%%%%%%%%%%%%%%%%%%%%%%%%%%%%%%%%%%%%%%%%%%%%%%%%%%%%%%%%%%%%%%%%%%%%%%%%%%

\section{Light scattering and N-photon imaging}

Having derived a description of the interaction between the atomic spins and the light field, we now consider how this interaction leads to light scattering and how the scattered light can be used to image the spin distribution. Taking the atomic spin quantization axis to be $\hat{\mathbf{z}}$, the spin operator in the $z$-direction is diagonal in atomic spin, that is $\hat{\rho}_{z}(x)=\sum_j\hat{\rho}_{z,j}|w(x-x_j)|^2$, where $\hat{\rho}_{z,j} =  \sum_{m}m\hat{n}_{j,m}$ and $\hat{n}_{j,m} = \hat{b}_{j,m}^\dagger\hat{b}_{j,m}$.
The atomic evolution in the optical lattice due to intersite tunneling and atom-atom interactions occurs on a time scale of order the inverse recoil frequency \cite{Jaksch1998a}. We will assume that the duration of the light matter interaction is short compared to this time scale and neglect the atomic evolution during the interaction. First order perturbation then mixes the original state of the atom-photon system $|\psi\rangle$ with
\begin{multline}
\int dx\int^{k}_{-k}d \kappa_1 d\kappa_2 
e^{i (\kappa_2-\kappa_1)x}\hat{\rho}_{z}(x)\\
\times\left(\frac{\gamma(\kappa_2)}{\gamma(\kappa_1)}\hat{a}^\dagger_1(\kappa_1)\hat{a}_2(\kappa_2)-
\frac{\gamma(\kappa_1)}{\gamma(\kappa_2)}\hat{a}^\dagger_2(\kappa_1)\hat{a}_1(\kappa_2)\right)|\psi\rangle
\label{first_order_scatter}
\end{multline}

This first order scattering process will transform the initial momentum distribution of the photons $\phi_i(\kappa_1,\ldots,\kappa_N)$ into the scattered distribution $\phi_s(\kappa_1,\ldots,\kappa_N)$.
The imaging system in Figure \ref{imaging_setup} will then map the scattered momentum distribution to a new momentum distribution at the detector $\phi_d(\kappa_1,\ldots,\kappa_N)$. Due to the finite extent of the lenses in the imaging system, the transverse momentum of the photons reaching the detector will be restricted to a finite bandwidth, such that $|\kappa|< \kappa_l\equiv k\sin\theta$ where $\sin\theta$ is the numerical aperture of the optical system \cite{Tsang2009a,Kolobov2000a}. This limit, which corresponds to the Rayleigh-Abbe diffraction limit, requires
\begin{equation}
\phi_d(\kappa_1,\ldots,\kappa_N) = \left\{ \begin{array}{ll} 0 &\text{if any }|\kappa_n|> \kappa_l \\
\phi_s(\kappa_1,\ldots,\kappa_N)  &\text{otherwise}
\end{array}\right.
\end{equation}  
where for simplicity we have let $f_1=f_2$. 

We now have a description of the $N$-photon state at the detector and we are left with the problem of extracting useful information from the state. A measurement of the intensity of the light at the detector provides the simplest form of image, but does not make full use of the multiphoton state. Using an $N$-photon absorbing material that creates a signal when all $N$ photons arrive at the same point allows us to make better use of the $N$-photon nature, but may be subject to low efficiency \cite{Tsang2007a}. Recently it was demonstrated in \cite{Tsang2009a} that intensity measurements at the detector followed by post-processing can reveal similar information to $N$-photon absorption. As this method does not rely on the photons all being coincident at a single point it has much higher efficiency.

The variable required in these measurements is the $N$-photon intensity distribution, $I(x_1,\ldots,x_N) = \sum_{\sigma_1,\ldots,\sigma_N}\langle \tilde{E}_{\sigma_N}^-(x_N)\ldots\tilde{E}_{\sigma_1}^-(x_1)\tilde{E}_{\sigma_1}^+(x_1)\ldots\tilde{E}_{\sigma_N}^+(x_N)\rangle$.  At the detector for the state $|n\rangle_x|N-n\rangle_y$ this is given by
% the sum of coordinate permutations of the $N$-dimensional Fourier transform of the momentum distribution
\begin{equation}
I(x_1,\ldots,x_N)
=\sum_{%\{x_1',\ldots,x_N'\}=
P}\eta^N\left|\psi_d(P(x_1,\ldots,x_N))\right|^2
\end{equation}
where the sum is over all coordinate permutations $P$ and
\begin{multline}
\psi_d(x_1,\ldots,x_N)\equiv\int_{-k}^{k} d\kappa_1\ldots d\kappa_N
\gamma(\kappa_1)\ldots\gamma(\kappa_N)
 \\\times\frac{1}{(2 \pi)^{N/2}}\exp\left(i \sum_{n=1}^N\kappa_n x_n\right)\phi_d(\kappa_1,\ldots,\kappa_N).
\end{multline}
is a $N$-dimensional transform of the momentum distribution.
The transform $\psi_i(x_1,\ldots,x_N)$ of the input light is defined in the same way.
Note that if all photons are in the same polarization state, the momentum distribution must obey bosonic symmetry so that $\phi(\ldots,\kappa_n,\ldots,\kappa_m,\ldots)=\phi(\ldots,\kappa_m,\ldots,\kappa_n,\ldots)$ for all $n$ and $m$. In which case $\psi(x_1,\ldots,x_N)$ must obey the same symmetry and is invariant under coordinate permutations.

%%%%%%%%%%%%%%%%%%%%%%%%%%%%%%%%%%%%%%%%%%%%%%%%%%%%%%%%%%%%%%%%%%%%%%%%%%%%%%%%%%%%%%%%%%%%%%%%%%%%%%%%%%%%%%%%%%%%%%%%%%%%%%%%%%%%%%%%%%%%%%%%%%%%%%%%%%%%%%%%%%%%%%%%%%%%%%%%%%%%%%%%%%%%%%%%%%%%%%%%%%%%%%%%%%%%%%%%%%%%%%%%%%%%%%%%%%%%%%%%%%%%%%%%%%%%%%%%

\section{Coherent spin imaging}

Before studying multiphoton imaging, we first look at the image created due to coherent laser photons scattering off the atoms in the optical lattice. We suppose the photons enter the system with $\boldsymbol{\epsilon_2}$-polarization and momentum distribution $\phi_i(\kappa)$. First order scattering according to Equation (\ref{first_order_scatter})
will then result in $\boldsymbol{\epsilon_1}$-polarized photons with a momentum distribution that depends on the atomic field operators according to
\begin{equation}
\phi_s(\kappa) = \frac{1}{\gamma(\kappa)}\int^k_{-k} d\kappa' \int dx \gamma(\kappa')e^{i (\kappa'-\kappa)x}\phi_i(\kappa')\hat{\rho}_z(x).
\end{equation}
Here, because light scattering off the atoms results in an orthogonal polarization, the unscattered light can be filtered out and we can detect an image due only to the scattered light \cite{Vega2008a}.
The intensity distribution at the detector will then be
\begin{equation}
I(x) = \eta\langle\psi_d^\dagger(x)\psi_d(x)\rangle
\label{single_photon_image}
\end{equation}
where
\begin{align}
\psi_d(x)&= \frac{1}{\sqrt{2 \pi}}\int_{-\kappa_l}^{\kappa_l} d\kappa \gamma(\kappa)\phi_s(\kappa)e^{i \kappa x} \nonumber \\
&= 2 \kappa_l \int dx'\sinc(\kappa_l(x-x'))\psi_i(x')\hat{\rho}_z(x').
\label{single_photon_amplitude}
\end{align}
The result is a diffraction limited image of the spin density weighted by the photons original spatial distribution. The integration over the $\sinc$ function in the image amplitude $\psi_d(x)$ blurs the image, but has the useful side effect that the image depends on the spin correlation function $\langle\hat{\rho}_z(x')\hat{\rho}_z(x'')\rangle$.

%%%%%%%%%%%%%%%%%%%%%%%%%%%%%%%%%%%%%%%%%%%%%%%%%%%%%%%%%%%%%%%%%%%%%%%%%%%%%%%%%%%%%%%%%%%%%%%%%%%%%%%%%%%%%%%%%%%%%%%%%%%%%%%%%%%%%%%%%%%%%%%%%%%%%%%%%%%%%%%%%%%%%%%%%%%%%%%%%%%%%%%%%%%%%%%%%%%%%%%%%%%%%%%%%%%%%%%%%%%%%%%%%%%%%%%%%%%%%%%%%%%%%%%%%%%%%%%%

\section{Two-photon spin image}

We next consider the scattering of a two-photon state from the atoms in the optical lattice and the resulting two-photon intensity distribution. We take the initial state to have one $\boldsymbol{\epsilon_1}$-polarized photon and one $\boldsymbol{\epsilon_2}$-polarized photon, then first order scattering from this state will result in a state with two photons with the same polarization. The photons in the final state will then be indistinguishable and knowledge of which photon was scattered is lost. Using polarization filters or beam splitters, we can then detect only two-photon states with two photons in the same polarization state. The resulting image is then due to pairs of photons containing one scattered photon and one unscattered photon.
If we filter out $\boldsymbol{\epsilon_2}$-polarized photons, the two-photon intensity distribution is
\begin{align}
I(x_1,x_2) &= \langle \tilde{E}_1^+(x_1)\tilde{E}_1^+(x_2)\tilde{E}_1^-(x_2)\tilde{E}_1^-(x_1)\rangle\nonumber\\
&= 2\eta^2 \langle \psi_d^\dagger(x_1,x_2)\psi_d(x_1,x_2)\rangle
\end{align}
where
\begin{multline}
\psi_d(x_1,x_2) =\sqrt{2} \kappa_l \int dx \hat{\rho}_z(x) \left(\sinc(\kappa_l(x_1-x))\psi_i(x_2,x)\right.\\\left.
+\sinc(\kappa_l(x_2-x))\psi_i(x_1,x)\right).
\label{two_photon_amp}
\end{multline}
The amplitude $\psi_d(x_1,x_2)$ consists of two terms due to the two indistinguishable paths the photons could have taken to reach $x_1$ and $x_2$, that is, the scattered photon reaching $x_1$ and the unscattered photon reaching $x_2$ and \textit{vice versa}. For simplicity here we have assumed that the photon source is also limited in transverse momentum by $\kappa_l$.

The use of two-photon absorption imaging can now lead to higher resolution than that of the coherent spin image discussed in the previous section. If the initial two-photon state is generated by parametric down conversion, then the transverse momenta of the photons is typically anti-correlated \cite{Klyshko1988a}. If we take the initial momentum distribution to be approximately $\phi_i(\kappa_1,\kappa_2) = \delta(\kappa_1+\kappa_2)$ for $|\kappa_1|,|\kappa_2|<\kappa_l$ and zero elsewhere, then the spatial distribution is $\psi_i(x,x')= \frac{\kappa_l}{\pi}\sinc(\kappa_l(x-x'))$. In this case a two-photon absorption image depends on the amplitude
\begin{equation}
\psi_d(x_1,x_1) = \frac{\sqrt{8} \kappa_l^2}{\pi}\int dx \hat{\rho}_z(x)\sinc^2(\kappa_l(x_1-x)).
\end{equation}
Comparing this with the coherent imaging amplitude in Equation (\ref{single_photon_amplitude}), we see that using two photons here results in the $\sinc$ function being raised to the power of two, allowing an image with higher resolution. In general for an $N$-photon state this type of imaging results in an improvement in resolution that scales like $\sqrt{N}$ \cite{Giovannetti2009a}.

A two-photon absorption image requires two photon detectors. A more efficient way to acquire the resolution improvement is to calculate the centroid distribution of the two-photon intensity distribution \cite{Tsang2009a}.
In place of the Cartesian coordinates we can express the system in terms of the centroid and relative positions of the two photons
\begin{equation}
X \equiv \frac{x_1+x_2}{2},		\qquad		\xi \equiv \frac{x_1-x_2}{2}.
\end{equation}
For the parametric down converted source described above the centroid probability distribution (unnormalized) is
\begin{multline}
\int d\xi\langle\psi_d^\dagger(X+\xi,X-\xi)\psi_d(X+\xi,X-\xi)\rangle = \\
\frac{8( \kappa_l)^4}{\pi^2}\int dx dx' \langle\hat{\rho}_z(x)\hat{\rho}_z(x')\rangle
K(X,x,x'),\label{centroid_int}
\end{multline}
where
\begin{multline}
K(X,x,x') = \int d\xi \sinc(\kappa_l(X+\xi-x)) \sinc(\kappa_l(X-\xi-x))\\
\times \sinc(\kappa_l(X+\xi-x')) \sinc(\kappa_l(X-\xi-x')).
\end{multline}
This results in a similar resolution improvement as for the two-photon absorption above, but is expected to be much simpler to implement in an experiment.

If we could measure the amplitude $\langle\psi(x_1,x_2)\rangle$ directly, for example by spatial quantum state tomography \cite{Raymer2009a}, then the centroid measurement applied to the amplitude would result in an image with higher resolution still.
Using the identity
\begin{equation}
\int^\infty_{-\infty} dx \sinc(b(x+y)) \sinc(b(z-x)) = \frac{\pi}{b} \sinc(b(y+z))
\end{equation}
the probability distribution (unnormalized) of the centroid becomes
\begin{multline}
\left|\int d\xi \langle\psi(X+\xi,X-\xi)\rangle\right|^2 = \\
8\kappa_l^2\left|\int dx \langle\hat{\rho}_z(x)\rangle \sinc(2 \kappa_l (X - x))\right|^2.
\label{centroid_amp}
\end{multline}
The factor of two is now inside the $\sinc$ function, leading to a factor of two resolution improvement over the coherent case. However, the requirement for quantum state tomography means that this would be significantly more difficult to achieve in an experiment.

%%%%%%%%%%%%%%%%%%%%%%%% FIGURE 2 %%%%%%%%%%%%%%%%%%%%%%%%%%%%%%%%%

\begin{figure}
\begin{center}
\includegraphics{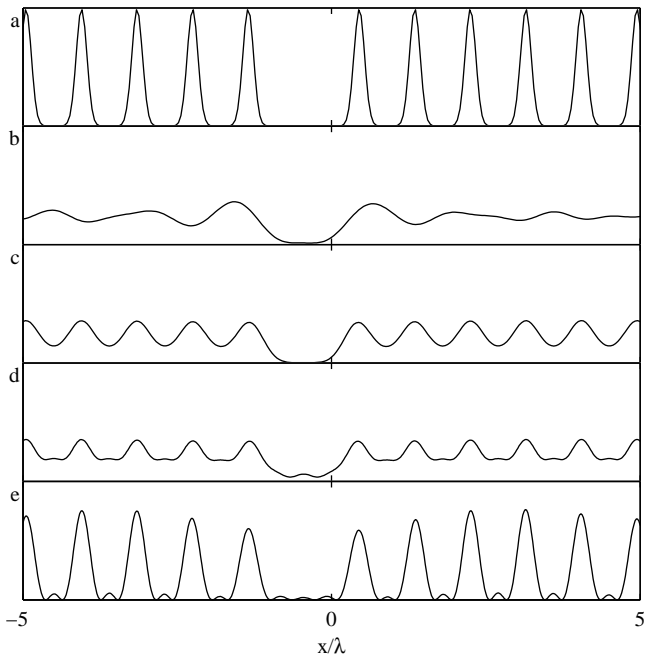}
\end{center}
\caption{Normalized images of a lattice containing atoms all in the $m = 1$ state with a defect. a) The lattice spin density to be imaged. b) The coherent image. c) The two-photon absorption image. d) The centroid distribution for the two-photon intensity distribution (Eq. (\ref{centroid_int})). e) The centroid of the two-photon amplitude (Eq. (\ref{centroid_amp})). Here the lattice site separation is $0.9\lambda$ and $\sin\theta = 2/3$.}
\label{imaging_comparison}
%image created by compare_spin_quantum_image.m
\end{figure}

%%%%%%%%%%%%%%%%%%%%%%%% FIGURE 2 %%%%%%%%%%%%%%%%%%%%%%%%%%%%%%%%%

In Figure \ref{imaging_comparison} we compare the various possible images described above. The images are of an optical lattice with a single $m=1$ atom at each site except for the missing or $m=0$ atom at $x=-0.45 \lambda$. The lattice site separation is $0.9$ times the wavelength of the imaging light and the optical system has a numerical aperture of 2/3. For these parameters the coherent image allows the identification of the defect in the optical lattice filling, but does not resolve the individual sites. The two-photon absorption image and the centroid of the two-photon intensity distribution both resolve the individual sites. The centroid of the two-photon amplitude resolves the sites with an improved visibility. 

The precise resolving power of each setup depends on the lattice properties, but a rough guide it that in the coherent image it becomes difficult to resolve individual sites for site spacings below $a\sim\lambda/\sin\theta$, while for the two-photon amplitude centroid this limit is halved. For the two-photon absorption and two-photon intensity centroid images individual peaks associated with each lattice also develop around $a\sim\lambda/2\sin\theta$ as in the amplitude centroid image, but the visibility of these peaks is lower than for the amplitude centroid image.

%%%%%%%%%%%%%%%%%%%%%%%%%%%%%%%%%%%%%%%%%%%%%%%%%%%%%%%%%%%%%%%%%%%%%%%%%%%%%%%%%%%%%%%%%%%%%%%%%%%%%%%%%%%%%%%%%%%%%%%%%%%%%%%%%%%%%%%%%%%%%%%%%%%%%%%%%%%%%%%%%%%%%%%%%%%%%%%%%%%%%%%%%%%%%%%%%%%%%%%%%%%%%%%%%%%%%%%%%%%%%%%%%%%%%%%%%%%%%%%%%%%%%%%%%%%%%%%%

\section{Images of example spin states}

We now apply our imaging proposal to example spin states of atoms in an optical lattice.
For an optical lattice with a single spin-1 atom at each site, the ground state potentially consists of dimers \cite{Yip2003a}. This occurs when neighboring sites form a spin singlet (total spin equal to zero). Denoting site $j$'s spin states $m=+1,0,-1$ by $|+\rangle_j$,$|0\rangle_j$ and $|-\rangle_j$ respectively, a dimerized state of an $M$-site lattice ($M$ even) is represented by
\begin{equation}
\bigotimes^{M/2}_{j=1}\frac{1}{\sqrt{3}}\left(|+\rangle_{2j}|-\rangle_{2j-1}+|-\rangle_{2j}|+\rangle_{2j-1}-|0\rangle_{2j}|0\rangle_{2j-1}\right).
\end{equation}

The coherent imaging intensity can be rewritten in terms of the site correlation functions as
\begin{equation}
I(x) = 4 \eta\kappa_l^2\sum_{j,r}  \langle\hat{\rho}_{z,j}\hat{\rho}_{z,r}\rangle f(x-x_j)f(x-x_r)
\end{equation}
where
\begin{equation}
f(x) = \int dx'|w(x') |^2 \sinc(\kappa_l(x-x')),
\end{equation}
and $|w(x)|^2 \equiv \int dy dz|w(x,y,z)|^2$. For the dimer case $\langle \hat{\rho}_{z,j}^2\rangle = 2/3$, 
$\langle \hat{\rho}_{z,2j-1} \hat{\rho}_{z,2j}\rangle = -2/3$ for integer $j$ and $\langle \hat{\rho}_{z,j} \hat{\rho}_{z,r}\rangle =0 $ otherwise.
The intensity then becomes
\begin{equation}
I(x) = \frac{8\eta\kappa_l^2}{3} \sum_{j=1}^{M/2}  \left(f(x-x_{2j-1}) - f(x-x_{2j})\right)^2.
\end{equation}

From this form we see that the atomic correlations lead to interference in the resulting image. In particular, at the centre of each dimer pair, $x = (x_{2j-1}+x_{2j})/2$, the contribution to the image from the atoms in the dimmer at $x_{2j-1}$ and $x_{2j}$ cancel, and we are left with a background intensity due to the other atoms. This cancellation leads to local minima in the image at the center of each dimer pair. The cancellation does not occur at $x = (x_{2j}+x_{2j+1})/2$ between atoms from separate dimers. For a lattice spacing $a \le 1/(2 \sin\theta)$ the images of sites $2j$ and $2j+1$ are no longer resolved and merge at $x = (x_{2j}+x_{2j+1})/2$ to form local maxima. Thus below the diffraction limit, the image is an oscillation with double the spatial period of the lattice due to the paring of atoms. This pattern persists until oscillations with double the lattice period can no longer be resolved.

Another potential state of the spin-1 chain is a trimerized state, where three neighboring sites form a singlet \cite{Nomura1991a}. A trimerized state of an $M$-site lattice ($M$ multiple of three) is
\begin{multline}
\bigotimes^{M/3}_{j=1}\frac{1}{\sqrt{6}}\left(|+0-\rangle_j+|-+0\rangle_j+|0-+\rangle_j\right.
\\\left.-|+-0\rangle_j-|0+-\rangle_j-|-0+\rangle_j\right)
\end{multline}
where $|m_1 m_2 m_2\rangle_j=|m_1\rangle_{3j-2}|m_2\rangle_{3j-1}|m_3\rangle_{3j}$. In this case $\langle \hat{\rho}_{z,j}^2\rangle = 2/3$, 
$\langle \hat{\rho}_{z,3j-2} \hat{\rho}_{z,3j-1}\rangle=\langle \hat{\rho}_{z,3j-2} \hat{\rho}_{z,3j}\rangle=\langle \hat{\rho}_{z,3j-1} \hat{\rho}_{z,3j}\rangle = -2/6$ for integer $j$ and $\langle \hat{\rho}_{z,j} \hat{\rho}_{z,r}\rangle =0 $ otherwise. 

The coherent image intensity then becomes
\begin{multline}
I(x) = \frac{4\eta\kappa_l^2}{3} \sum_{j=1}^{M/3} \left( (f(x-x_{3j-2}) -f(x-x_{3j-1}))^2\right.\\
+(f(x-x_{3j-2}) - f(x-x_{3j}))^2\\\left.
+(f(x-x_{3j-1}) - f(x-x_{3j}))^2\right).
\end{multline}
As in the case of the dimer, the atomic correlations lead to interference in the image. In this case the interference results in an oscillation with three times the spatial period of the lattice, with local minima at the center of each trimer and maxima at the points between separate trimers.

\begin{figure}
\begin{center}
\includegraphics{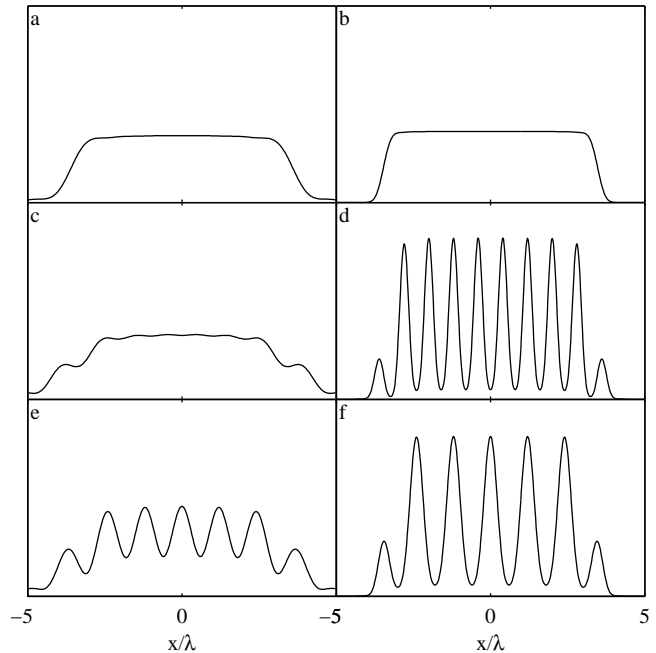}
\end{center}
\caption{Normalized coherent (left column) and two-photon intensity centroid (right column) images of an 18 atom spin chain in the
(a-b) unpolarized state , (c-d) dimer state and (e-f) trimer state. The images are for a lattice spacing $0.4\lambda$ and $\sin\theta = 1/2$.}
\label{dimer_trimer_sp_tp}
%image created by lp_dimer_trimer_compare_sp_and_tp_image.m
\end{figure}

The same effects occur in the two-photon images of both the dimer and trimer, but due to the greater resolving power they are visible for smaller lattice site spacings. In Figure \ref{dimer_trimer_sp_tp} the coherent and two-photon intensity centroid images of the the dimer and trimer states are compared with the unpolarized state that has $\langle\hat{\rho}_{z,j}\rangle = 0$, $\langle \hat{\rho}_{z,j}^2\rangle = 2/3$ and $\langle \hat{\rho}_{z,j} \hat{\rho}_{z,r}\rangle = 0$ for $j\neq r$.
The images are generated for parameters where the dimer oscillations become difficult to resolve in the coherent image.
The dimer oscillations in the coherent image become unresolvable for lattice spacings below $a\sim 0.2\lambda/\sin\theta$, whereas the oscillations in the two photon image become unresolvable at $a \sim 0.14\lambda/\sin\theta$. The oscillations in the trimer state become unresolved in the coherent image below $a\sim 0.16\lambda/\sin\theta$, and in the two-photon image below $a\sim 0.1\lambda/\sin\theta$.

%%%%%%%%%%%%%%%%%% FIGURE 3 %%%%%%%%%%%%%%%%%%%%%%%%%%%%%%%%%%%

%%%%%%%%%%%%%%%%%% FIGURE 3 %%%%%%%%%%%%%%%%%%%%%%%%%%%%%%%%%%%

%%%%%%%%%%%%%%%%%%%%%%%%%%%%%%%%%%%%%%%%%%%%%%%%%%%%%%%%%%%%%%%%%%%%%%%%%%%%%%%%%%%%%%%%%%%%%%%%%%%%%%%%%%%%%%%%%%%%%%%%%%%%%%%%%%%%%%%%%%%%%%%%%%%%%%%%%%%%%%%%%%%%%%%%%%%%%%%%%%%%%%%%%%%%%%%%%%%%%%%%%%%%%%%%%%%%%%%%%%%%%%%%%%%%%%%%%%%%%%%%%%%%%%%%%%%%%%%%

\section{Spatial correlation measurements}

Another potential use of two photon imaging is to examine local spatial correlations in the atomic lattice that cannot be accessed using methods that probe the entire sample \cite{Eckert2008a,Roscilde2009a}. If we have light that is spatial correlated as $\phi_i(x,x')= \frac{\kappa_l}{\sqrt{2}\pi}(\sinc(\kappa_l(x-x'-a))+\sinc(\kappa_l(x-x'+a)))$, so that the two photons are separated by $a$, then the two-photon image amplitude (Equation (\ref{two_photon_amp})) becomes
\begin{multline}
\psi_d(x_1,x_2) = \frac{ \kappa_l^2}{\pi}\int dx (\hat{\rho}_z(x)+\hat{\rho}_z(x+a))\\
\times(\sinc(\kappa_l(x_1-x))\sinc(\kappa_l(x_2-x-a))+\\
\sinc(\kappa_l(x_1-x-a))\sinc(\kappa_l(x_2-x)).
\end{multline}
A measurement of the two-photon intensity distribution at $x_1 = x $ and $ x_2 = x+a$ then depends on $\langle(\hat{\rho}_z(x)+\hat{\rho}_z(x+a))^2\rangle$. For an optical lattice, if the two photon resolution is such that 
$\int dx |w(x)|^2\sinc^2(\kappa_l x)\gg\int dx |w(x)|^2\sinc^2(\kappa_l (x-a))$
then the main contribution to the two-photon distribution at $(x_1,x_2) = (x_j,x_{j+1})$ is
\begin{multline}
\langle\psi^\dagger_d(x_j,x_{j+1})\psi_d(x_j,x_{j+1})\rangle \approx \frac{ \kappa_l^4}{\pi^2}\langle(\hat{\rho}_{z,j}+\hat{\rho}_{z,j+1})^2\rangle
\\\times\left(\int dx |w(x)|^2\sinc^2(\kappa_l x)\right)^2.
\end{multline}
This allows us to access the neighbor correlation using $\langle(\hat{\rho}_{z,j}+\hat{\rho}_{z,j+1})^2\rangle = \langle\hat{\rho}_{z,j}^2+\hat{\rho}_{z,j+1}^2\rangle + 2 \langle\hat{\rho}_{z,j}\hat{\rho}_{z,j+1}\rangle$.  For the dimer state the atoms of a dimer pair have a neighbor correlation leading to $I(x_{2j-1},x_{2j}) \approx \langle\rho_{2j-1}^2+\rho_{2j}^2\rangle + 2\langle\rho_{2j-1}\rho_{2j}\rangle=0$.  There is no neighbor correlation between atoms of separate dimers leading to $I(x_{2j},x_{2j+1}) \approx \langle\hat{\rho}_{z,2j}^2+\hat{\rho}_{z,2j+1}^2\rangle =4/3$. 
For the trimer state the correlation between neighbors in a trimer leads to $I(x_{3j-2},x_{3j-1})=I(x_{3j-2},x_{3j})=I(x_{3j-1},x_{3j}) \approx 2/3$. While atoms of separate trimers have no correlation leading to $I(x_{3j},x_{3j+1}) \approx  4/3$

\begin{figure*}
\begin{center}
\includegraphics{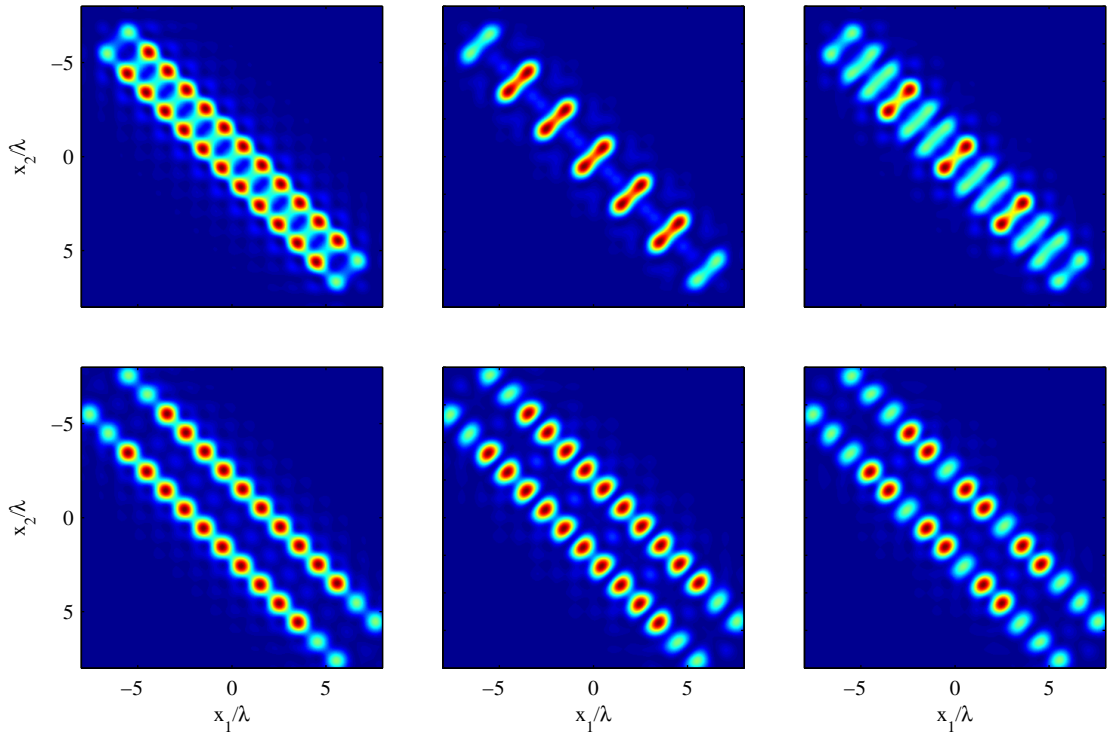}
\end{center}
\caption{Images of two-photon intensity for spatial correlation measurement using photons initially correlated to be separated by $a$ (top row) and $2a$ (bottom row). Left column: unpolarized state. Middle column: dimer state. Right column: trimer state. The images are for $\sin\theta = 1/2$ and $a = \lambda$.}
\label{spatial_correlations}
%image created by lp_dimer_trimer_compare_corr_image.m
\end{figure*}

In the first row of Figure \ref{spatial_correlations} we compare the resulting two-photon images for the trimer and dimer states with the unpolarized state. The correlation between atoms in the dimer state leads to the removal of every second peak compared to the unpolarized state distribution, while the trimer correlations lead to partial cancellation of the peaks corresponding to the correlated atoms in each trimer.

Similarly, we could also consider light correlated so that photons in the initial pair were separated by $2a$. The two-photon image would then probe the correlation $\langle \hat{\rho}_{z,j} \hat{\rho}_{z,j+2}\rangle$. For the dimer no such correlations exist and cancellation does not occur as it did in the previous example. The trimer does have correlations over this range and partial cancellation of peaks occurs. This behavior is shown in the second row of images in Figure \ref{spatial_correlations}.

%%%%%%%%%%%%%%%%%%%%%%%%%%%%%%%%%%%%%%%%%%%%%%%%%%%%%%%%%%%%%%%%%%%%%%%%%%%%%%%%%%%%%%%%%%%%%%%%%%%%%%%%%%%%%%%%%%%%%%%%%%%%%%%%%%%%%%%%%%%%%%%%%%%%%%%%%%%%%%%%%%%%%%%%%%%%%%%%%%%%%%%%%%%%%%%%%%%%%%%%%%%%%%%%%%%%%%%%%%%%%%%%%%%%%%%%%%%%%%%%%%%%%%%%%%%%%%%%

\section{Conclusion}

As experiments with ultracold atoms in optical lattices become more complex, there is a need for methods of probing the resulting states that provide more detailed information. Here we have demonstrated how coherent \textit{in situ} imaging can distinguish between spin states in an optical lattice. Further we have shown how the use of spatially correlated light can improve the resolution of these images and allow us to distinguish spin states beyond the limits imposed by Rayleigh diffraction. We have also demonstrated a method that uses spatially correlated light to directly probe the local spatial correlation function of the atoms. While we have restricted most of our attention here to two-photon imaging due to the relative ease of generating two-photon states in experiments, these methods can be readily generalized to higher photon number states.

%%%%%%%%%%%%%%%%%%%%%%%%%%%%%%%%%%%%%%%%%%%%%%%%%%%%%%%%%%%%%%%%%%%%%%%%%%%%%%%%%%%%%%%%%%%%%%%%%%%%%%%%%%%%%%%%%%%%%%%%%%%%%%%%%%%%%%%%%%%%%%%%%%%%%%%%%%%%%%%%%%%%%%%%%%%%%%%%%%%%%%%%%%%%%%%%%%%%%%%%%%%%%%%%%%%%%%%%%%%%%%%%%%%%%%%%%%%%%%%%%%%%%%%%%%%%%%%%

\bibliography{../bib/dphil}

%%%%%%%%%%%%%%%%%%%%%%%%%%%%%%%%%%%%%%%%%%%%%%%%%%%%%%%%%%%%%%%%%%%%%%%%%%%%%%%%%%%%%%%%%%%%%%%%%%%%%%%%%%%%%%%%%%%%%%%%%%%%%%%%%%%%%%%%%%%%%%%%%%%%%%%%%%%%%%%%%%%%%%%%%%%%%%%%%%%%%%%%%%%%%%%%%%%%%%%%%%%%%%%%%%%%%%%%%%%%%%%%%%%%%%%%%%%%%%%%%%%%%%%%%%%%%%%%

\end{document}